\documentclass[mathleft
]{an}
\usepackage{graphicx,lscape}
\usepackage{times}
\usepackage{appendix}

\begin{document}

\Pagespan{789}{}
\Yearpublication{2006}%
\Yearsubmission{2005}%
\Month{11}%
\Volume{999}%
\Issue{88}%

\title{Multiperiodic semiregular variable stars in the ASAS data base:\\ A pilot study
}

\author{I. Fuentes-Morales\thanks{Corresponding author:
  \email{irma.fuentes@uv.cl}\newline}
\&  N. Vogt
}
\titlerunning{Multiperiodic semiregular variable stars in the ASAS data base}
\authorrunning{I. Fuentes-Morales \& N. Vogt}
\institute{
Instituto de F\'{i}sica y Astronom\'{i}a , Universidad de Valpara\'{i}so, 
Avenida Gran Breta\~na 1111, Valpara\'{i}so, Chile}

\received{-}
\accepted{-}
\publonline{later}



\abstract{Based on a search for multi-periodic variability among the semi-regular red variable stars in the database of the All Sky Automated Survey (ASAS), a sample of 72 typical examples is presented. Their period analysis was performed using Discrete Fourier Transform.  In 41 stars we identified two significant periods each, simultaneously present, while the remaining 31 cases revealed even three such periods per star.  They occur in a range roughly between 50 and 3000 days. Inter-relationships between these periods were analyzed using the ``double period diagram'' which compares adjacent periods, and the so-called ``Petersen diagram'', the period ratio vs. the shorter period. In both diagrams we could identify six sequences of accumulation of the period values.  For five of these sequences (containing 97\% of all data points) we found an almost perfect coincidence with those of previous studies which were based on very different samples of semi-regular red variables.  Therefore, existence and locations of these sequences in the diagrams seem to be universal features, which appear in any data set of semi-regularly variable red giants of the AGB; we conclude that they are caused by different pulsation modes as the typical and consistent properties of similar stellar AGB configurations.  Stellar pulsations can be considered as the principal cause of the observed periodic variability of these stars, and not binary, rotation of a spotted surface or other possible reasons suggested in the literature. }

\maketitle

\section{Introduction}

The class of ``semiregular variables'' (SR) is quite common. The General Catalogue of Variable Stars, Moscow (GCVS; Samus et al. 2007, 2013\nocite{samus}) lists more than 6000 representatives, about 15\% of all variables in the GCVS. They are late K- or M type giants or supergiants located on the asymptotic giant branch (AGB) and display  quasiperiodic oscilations with periods between 30 and several thousand days, but with smaller amplitudes ($\leq$ 2.5 mag) than those of Mira stars. SR and Mira variables are primary sources for the enrichment of the interstellar medium with heavy elements, and therefore important for a proper understanding of the Milky Way evolution. On the other hand, their inner physical processes which cause their observed pulsations may contain crucial hints for unveiling the advanced evolution of late type giant stars.\\ 
\noindent
The physical understanding of this class is far from being satisfactory. Most authors agree that at least part of the periodic behaviour corresponds to stellar pulsation, but it is not clear whether the fundamental frequency or some of the overtones are excited. A rather common property is ``multiperiodicity'', i.e. the simultaneous presence of two, three or even more different periods in the same star. Often we can observe long periods from 500 up to several thousand days, combined with shorter ones in the range 50$-$400 days. Especially the long periods are difficult to reconcile with current pulsation theories. (Olivier \& Wood, 2005).  Therefore, alternative explanations for the long periods have been suggested: stellar rotation combined with surface irregularities (spots), binary companions or episodic dust ejection (Olivier \& Wood, 2003 and references therein).\\
There is a powerful tool in order to attack this uncertainty in the interpretation of the long periods: if one could establish that the observed period ratios in multiperiodic SR variables do obey certain patterns of preferred sequences in period vs. period diagrams (P vs. P), and if these sequences have universal validity, avoiding systematically other regions in the P vs. P diagrams, a strong argument in favour of pulsations in different modes would be obtained, as the only cause of SR variability. If any other possible physical explanation is acting, as those mentioned above, no significant groupings in the P vs. P diagrams would be expected. \\
\noindent
This idea is not new: Kiss et al. (1999)\nocite{kiss} studied a total of 93 bright semiregular variables, based on visual amateur observations obtained during several decades, and identified three certain and two less obvious sequences in the distribution of periods or period ratios in multiperiodic SR variables, analysing 44 stars with two and 12 stars with three significant and essentially stable periods. \\
The aim of our paper is to show that similar investigations can be carried out using the ``All Sky Automated Survey'' (ASAS), a data base of about 50000 variable stars of all kind, obtained at Las Campanas Observatory (LCO), Chile,  between 1997 and 2006 approximately (Pojmanski, 1997). In section 2 we describe briefly some details about ASAS and our target selection strategy and analysis methods; section 3 contains our results, which are discussed in Section 4, together with some final conclusions. 

\begin{table*}[ht!]
\centering

\caption{ Double periodic stars. ASAS name, other name, variability class (see text for details),  number of single observations per star in the ASAS data base $N_{s}$,  averaged V magnitude for each star $\overline{m}$, observation time covered $\Delta T$, Semi-amplitudes $A_{i}$ ($i$ = 0,1) and the longer and shorter periods $P_{i}$ ($i$ = 0,1) are given.}
\label{2P-tabla}
\resizebox {1\textwidth}{!} {
\begin{tabular}{cccccccccc}
\hline \hline 
\\
ASAS name    & Other name &Variability&$N_{s}$& $\overline{m}$  &$\Delta T$&$P_{0}$      &$A_{0}$     &$P_{1}$      & $A_{1}$   \\
& &class&&(mag)&(d)&(d)&(mag)&(d)&(mag)\\
\\

\hline \hline
054155+1450.1 & V535~Ori  &SRPV  & 456     &9.1  &2544     &726.9(3.4)   &0.39(0.01)  &82.3(0.2)   &0.09(0.02) \\ 
060431-3403.5 & AC~Col    &SRPV  & 659     &9.0  &3298     &349.0(0.6)   &0.48(0.01)  &66.9(0.3)   &0.05(0.02) \\ 
062312-7115.5 & -         &SRPV  & 505     &13.1 &2589     &761.5(11.1)  &0.26(0.02)  &75.5(0.1)   &0.19(0.02) \\ 
063833-7733.3 & -         &LSP   & 860     &12.8 &3297     &740.9(5.8)   &0.16(0.01)  &65.0(0.1)   &0.11(0.01) \\ 
065955-4232.0 & -         &LSP   & 1145    &12.4 &3300     &511.6(1.1)   &0.37(0.01)  &61.8(0.1)   &0.07(0.01) \\ 
070322-1330.0 & -         &LSP   & 563     &13.2 &3300     &750.0(3.4)   &0.47(0.01)  &136.4(0.8)  &0.10(0.02) \\ 
070853+0548.0 & -         &LSP   & 474     &13.4 &2616     &1275.5(14.3) &0.37(0.02)  &149.1(1.5)  &0.10(0.02) \\ 
072523-4041.7 & -         &RSG   & 659     &12.9 &3300     &804.9(5.3)   &0.33(0.01)  &90.7(0.1)   &0.21(0.02) \\ 
073127-0555.6 & -         &LSP   & 526     &12.4 &3300     &589.3(2.2)   &0.34(0.01)  &63.2(0.2)   &0.06(0.02) \\ 
074715-3612.6 & -         &LSP   & 661     &12.5 &3298     &1030.9(5.4)  &0.34(0.01)  &116.7(0.2)  &0.12(0.01) \\ 
081250-3133.2 & -         &LSP   & 618     &13.4 &3300     &1834.9(28.1) &0.32(0.01)  &178.4(0.4)  &0.16(0.01) \\ 
081735-5431.0 & -         &LSP   & 934     &10.9 &3300     &825.0(3.5)   &0.28(0.01)  &91.4(0.2)   &0.09(0.01) \\
082106-1256.9 & -         &LSP   & 748     &13.0 &3298     &634.2(3.0)   &0.33(0.01)  &64.5(0.1)   &0.16(0.02) \\
090734-5349.9 & -         &LSP   & 1034    &13.3 &3300     &545.5(4.7)   &0.15(0.02)  &66.9(0.1)   &0.12(0.01) \\
093941-2335.7 & AX~Hya    &LSP   & 594     &11.9 &3295     &615.9(2.9)   &0.31(0.01)  &57.8(0.1)   &0.13(0.02) \\ 
094336-7214.8 & -         &LSP   & 783     &13.4 &3181     &691.5(3.8)   &0.30(0.01)  &76.5(0.1)   &0.16(0.02) \\
100059-3458.7 &-          &SRPV  & 639     &12.2 &3298     &117.8(0.3)   &0.10(0.02)  &72.2(0.1)   &0.16(0.02) \\  
102149-3323.7 & -         &SRPV  & 700     &9.7  &3290     &671.4(2.6)   &0.29(0.01)  &84.1(0.1)   &0.13(0.01) \\
102551-4524.5 & -         &LSP   & 590     &13.5 &3298     &599.8(2.9)   &0.36(0.02)  &65.7(0.1)   &0.21(0.02) \\
103327-4601.8 & -         &SRPV  & 523     &13.6 &3294     &2531.6(110.3)&0.14(0.02)  &902.5(18.0) &0.14(0.02) \\
103910-7757.4 & EZ~Cha    &SRPV  & 1130    &9.6  &3297     &955.6(5.1)   &0.29(0.01)  &119.0(0.2)  &0.13(0.01) \\
110903-7240.0 & -         &LSP   & 1149    &13.3 &3297     &3878.8(308.4)&0.16(0.01)  &785.0(5.1)  &0.24(0.01) \\
111606-4011.7 & -         &LSP   & 627     &12.8 &3296     &549.3(1.4)   &0.45(0.01)  &66.7(0.3)   &0.08(0.02) \\
112612-5121.6 & NSV05201  &SRPV  & 804     &9.6  &3299     &1375.5(9.6)  &0.44(0.01)  &135.2(0.2)  &0.21(0.02) \\
130830-5923.1 & V0592~Cen &RCB   & 1239    &12.5 &3291     &1028.8(2.3)  &1.15(0.02)  &41.4(0.1)   &0.29(0.04) \\
131632-4442.3&UY~Cen      & SRPV & 660     &8.1  &3202     &1488.1(24.5) &0.39(0.03)  &180.4(0.4)  &0.29(0.03) \\   
133910-2035.9 & -         &SRPV  & 676     &12.0 &3186     &965.3(10.4)  &0.20(0.01)  &78.2(0.1)   &0.13(0.01) \\
135031-5957.1 & -         &LSP   & 877     &13.2 &3197     &1775.9(59.1) &0.14(0.02)  &555.9(1.7)  &0.37(0.01) \\
142927-5812.2 & -         &SRPV  & 650     &11.2 &3201     &1207.7(9.2)  &0.34(0.01)  &120.6(0.2)  &0.20(0.02) \\
150155-4354.7 & -         &SRPV  & 630     &13.0 &3178     &605.3(4.7)   &0.20(0.01)  &80.1(0.1)   &0.02(0.02) \\
150412-4811.3 & -         &LSP   & 582     &9.5  &3191     &348.7(0.6)   &0.31(0.01)  &48.1(0.2)   &0.04(0.02) \\
155856-5616.4 & -         &SRPV  & 410     &13.7 &3200     &1103.8(15.9) &0.29(0.03)  &143.5(0.6)  &0.19(0.03) \\
162648-0237.0 & V0707~Oph &LSP   & 424     &11.6 &3169     &688.8(3.0)   &0.30(0.01)  &87.2(0.4)   &0.07(0.02) \\
170133-4805.3 & -         &SRPV  & 554     &12.8 &3205     &821.7(12.3)  &0.14(0.01)  &80.0(0.1)   &0.21(0.01) \\
170512-6305.6 & -         &LSP   & 635     &12.8 &3206     &352.3(0.8)   &0.33(0.01)  &65.9(0.2)   &0.07(0.02) \\
173701-4651.8 & -         &LSP   & 657     &13.4 &3183     &539.4(2.7)   &0.30(0.02)  &63.3(0.2)   &0.09(0.02) \\
174719-1730.8 & -         &LSP   & 653     &13.3  &3190    &911.3(5.4)   &0.44(0.02)  &94.8(0.2)   &0.20(0.02) \\
175015-2223.7 & NSV9741   &SRPV  & 1129    &11.2&3198      &1421.2(9.5)  &0.34(0.01)  &130.8(0.2)  &0.17(0.01) \\
180955-2802.0 & -         &LSP   & 514     &13.4 &3173     &1322.4(17.4) &0.31(0.02)  &148.0(0.3)  &0.18(0.02) \\
192623-1526.1 & -         &SRPV  & 550     &13.6 &3181     &1383.1(34.9) &0.20(0.02)  &80.3(0.1)   &0.16(0.02) \\
202427-5336.5 & -         &LSP   & 1032    &12.4 &3295     &433.6(0.8)   &0.34(0.01)  &82.8(0.2)   &0.06(0.01) \\
\hline                                                                    
\end{tabular}                                                                           
}
\end{table*}

\begin{table*}[ht!]
\centering

\caption{ Triple periodic stars. Subscripts $i$ ($i$ = 0,1,2) for semi-amplitudes and periods are the same way as in Table \ref{2P-tabla} as well as the other columns.}
\label{3P-tabla}
\scalebox{0.82} {
\begin{tabular}{cccccccccccc}  
\hline \hline 
\\
ASAS name    & Other name &variability & $N_{s}$& $\overline{m}$  &$\Delta T$&$P_{0}$     & $A_{0}$     &$P_{1}$      & $A_{1}$   &$P_{2}$     & $A_{2}$  \\ 
& &class&&(mag)&(d)&(d)&(mag)&(d)&(mag)&(d)&(mag)\\
\\
                     
\hline \hline
003753-8237.2 &BV~984     & SRPV  & 1202   &  9.5  & 3300    & 880.0(5.9)   & 0.19(0.01) &149.0(0.3) &0.15(0.01) &87.0(0.1)  &0.14(0.01)\\                  
052321-0434.3 &FX~Ori     & LSP   & 680    &  10.0 & 3297    & 1100.4(7.9)  & 0.46(0.02) &573.4(4.0) &0.17(0.02) &115.7(0.6) &0.12(0.03)\\          
063547-1305.0 &NSV03042   & LSP   & 908    &  12.0 & 3300    & 741.6(2.1)   & 0.62(0.01) &381.5(14.9)&0.13(0.02) &89.1(0.2)  &0.10(0.02)\\          
070915+2217.7 &NSV03423   & SRPV  & 488    &  11.3 & 2544    & 716.7(3.8)   & 0.47(0.01) &498.8(2.2) &0.13(0.01) &96.0(0.3)  &0.08(0.02)\\          
074204-1155.2 &-          & LSP   & 678    &  13.1 & 3300    & 1200.0(45.7) & 0.02(0.01) &680.4(3.1) &0.34(0.02) &77.9(0.3)  &0.09(0.02)\\          
075217-0329.1 &-          & LSP   & 612    &  12.0 & 3300    & 1676.4(36.5) & 0.10(0.02) &947.5(5.4) &0.39(0.01) &186.3(0.3) &0.17(0.01)\\          
080433-2523.4 &IM~Pup     & LSP   & 565    &  10.6 & 3291    & 645.3(1.9)   & 0.40(0.01) &329.1(3.3) &0.10(0.02) &76.8(0.2)  &0.10(0.02)\\          
082008-2616.5 &NSV04011   & SRPV  & 703    &  11.4 & 3297    & 856.4(5.9)   & 0.26(0.01) &165.3(0.8) &0.09(0.01) &100.8(0.1) &0.18(0.01)\\          
091738-4118.9 &-          & LSP   & 678    &  11.9 & 3297    & 672.9(3.5)   & 0.28(0.01) &392.5(8.3) &0.12(0.02) &70.8(0.1)  &0.09(0.02)\\          
094941-6925.4 &-          & LSP   & 396    &  12.2 & 2306    & 2304.1(119.8)& 0.21(0.02) &569.3(3.9) &0.36(0.02) &82.2(0.2)  &0.15(0.02)\\          
095017-2240.5 &NSV04647   & LSP   & 995    &  12.8 & 3300    & 3142.8(146.4)& 0.17(0.01) &814.8(3.5) &0.37(0.01) &83.6(0.2)  &0.12(0.02)\\         
095358-7636.9 &-          & RCB   & 1028   &  11.8 & 3298    & 1243.8(4.3)  & 0.67(0.01) &100.5(0.2) &0.21(0.02) &50.2(0.1)  &0.17(0.02)\\              
102427-2401.4 &-          & SRPV  & 983    &  11.2 & 3299    & 461.4(0.8)   & 0.35(0.01) &341.9(3.5) &0.15(0.01) &62.5(0.2)  &0.05(0.01)\\        
111037-7007.0 &-          & LSP   & 974    &  13.5 & 3299    & 1404.5(24.8) & 0.10(0.01) &485.1(2.8) &0.20(0.01) &68.5(0.1)  &0.10(0.01)\\        
115628-3808.9 &-          & SRPV  & 611    &  12.9 & 3177    & 825.1(9.4)   & 0.18(0.02) &126.6(0.4) &0.13(0.02) &81.5(0.2)  &0.12(0.02)\\        
131157-5757.8 &-          & SRPV  & 1236   &  11.9 & 3291    & 587.7(2.3)   & 0.22(0.01) &188.1(0.4) &0.13(0.01) &110.8(0.1) &0.14(0.01)\\             
135923-6435.1 &KP~Cen     & SRPV  & 574    &  13.1 & 3181    & 1589.8(44.0) & 0.17(0.02) &119.6(0.3) &0.16(0.02) &76.6(0.2)  &0.09(0.02)\\        
141600-6153.9 &V0417~Cen  & SRPV  & 595    &  11.8 & 3181    & 1818.2(15.2) & 0.58(0.02) &548.4(6.4) &0.23(0.03) &397.6(4.3) &0.11(0.03)\\        
143647-5647.9 &-          & SRPV  & 685    &  13.0 & 3194    & 912.5(8.8)   & 0.19(0.01) &164.6(0.6) &0.11(0.01) &84.6(0.2)  &0.10(0.01)\\       
162232-5349.2 &-          & RSG   & 1098   &  12.3 & 3207    & 3558.7(35.2) & 0.82(0.01) &782.1(7.6) &0.30(0.03) &80.9(0.4)  &0.07(0.03)\\        
163318-6007.1 &-          & LSP   & 817    &  11.0 & 3200    & 444.4(0.8)   & 0.37(0.01) &222.2(0.8) &0.12(0.01) &53.1(0.2)  &0.06(0.02)\\       
163635-7307.0 &LW~Aps     & SRPV  & 979    &  13.7 & 3230    & 134.8(0.4)   & 0.18(0.02) &126.9(0.2) &0.26(0.02) &73.2(0.1)  &0.16(0.02)\\       
164512-2652.8 &AZ~Sco     & SRPV  & 762    &  12.6 & 3178    & 276.3(1.5)   & 0.15(0.01) &172.2(0.5) &0.12(0.01) &94.9(0.2)  &0.10(0.01)\\       
165004-0345.1 &-          & LSP   & 498    &  13.0 & 3170    & 1864.5(53.2) & 0.12(0.01) &469.6(2.3) &0.15(0.01) &59.0(0.1)  &0.09(0.01)\\        
165600-5435.9 &-          & LSP   & 621    &  11.4 & 3200    & 556.5(1.7)   & 0.34(0.01) &280.7(1.3) &0.10(0.02) &60.20(0.1) &0.09(0.02)\\       
170532+0216.8 &V2363~Oph  & LSP   & 438    &  9.9  & 3162    & 549.8(2.4)   & 0.28(0.01) &90.3(0.3)  &0.09(0.02) &56.5(0.2)  &0.06(0.02)\\       
170655+0351.1 &-          & LSP   & 411    &  12.5 & 3161    & 454.8(1.3)   & 0.36(0.01) &229.0(1.3) &0.12(0.02) &54.6(0.1)  &0.06(0.02)\\          
173504-6045.5 &-          & LSP   & 901    &  11.3 & 3188    & 490.4(0.9)   & 0.43(0.01) &249.0(1.9) &0.14(0.02) &53.3(0.1)  &0.07(0.02)\\       
190102-1055.0 &UY~Aql     & LSP   & 476    &  12.5 & 3180    & 451.0(1.0)   & 0.35(0.01) &227.9(1.2) &0.09(0.02) &160.2(0.9) &0.06(0.02)\\      
194714+0737.8 &V0458~Aql  & MIRA  & 323    &  13.3 & 2648    & 3530.6(197.2)& 0.46(0.04) &147.5(0.8) &0.24(0.05) &139.7(0.3) &0.50(0.04)\\         
200129-2540.9 &-          & LSP   & 708    &  11.8 & 3202    & 438.6(0.7)   & 0.34(0.01) &220.0(0.7) &0.10(0.01) &94.3(0.5)  &0.06(0.02)\\         
                                     
\hline
\end{tabular}
}
\end{table*}

\section{ Data analysis}
\subsection{Target selection from ASAS}

The ASAS is a long-term project dedicated to constant photometric monitoring of the whole available sky, covering more than $10^{7}$ stars brighter than 14 magnitude. The ultimate goal of the project is the detection and investigation of any kind of the photometric variability. ASAS has produced an extensive Variable Star Catalogue (ACVS) of the southern hemisphere (declination $<$ +280) with a total of 50124 entries. The majority of them are new discoveries (80\%, as compared with GCVS). \\
\noindent
Our study is based on the ASAS-3 (ACVS) catalogue with photometric V-band data. On average about 500 data points per star have been observed in the course of nine years (1997 - 2006), with a typical photometric accuracy of $\pm$ 0.03 mag for stars brighter than 12 mag in V. For each star, a light curve can be displayed immediately using the ASAS web page \footnote{www.astrouw.edu.pl/asas} which also permits to see a phased light curve folded with the period suggested by ASAS (often not very reliable) or with any other period chosen by the user. The ASAS data base also gives a possible classification for each variable, if they could be considered as eclipsing binaries or as some class of pulsating stars (Cepheids, RR Lyr, $\delta$ Sct, $\beta$ Cep  or Mira). However, this classification comprises less than 40\% of all variables detected by ASAS. The remaining 30993 stars are just marked as class ``MISC'' (miscellaneous). Apparently the classification scheme of ASAS was unable to assign a class in these cases. Richards et al. (2012)\nocite{rich} presented a substantial improvement of this unsatisfactory situation: They developed machine-learning classification software for 28 different variability classes, and applied it to all 50124 variables in the ASAS data base. They list the resulting class for each star, together with a probability measure of this classification (and of other less probable classes, which could be assigned to it), as well as a period, which turns out to be more reliable than that given in the ASAS data base. \\
\noindent
In the course of another research project, we inspected the light curves of all ASAS variables classified ``MISC'' with amplitudes $\geq$ 1.0 mag (about 2600 targets in total), in order to check whether there are periodic variations among them. In fact, a large majority of them turned out to display long period variations ($\geq$ 60 days), immediately visible on the screen of the ASAS web page, and confirmed, in most cases, with phased light curves using the period listed by Richards et al. (2012)\nocite{rich}. In some cases, we could identify in the original ASAS light curve double periodicity:  a more rapid variation with periods in a typical range between 60 and 100 days, superimposed on a longer-term variability (500 $-$ 1000 days and more). These are the targets selected for the present paper.

\subsection{Time series analysis}

We used {\sc Period04} (Lenz \& Breger 2005 \nocite{lenz}) software to perform the period analysis which is based on Discrete Fourier Transform (DFT), and especially dedicated to statistical an-alysis of astronomical time series containing long gaps in the observational data. The Fourier analysis consists in a search for sinusoidal functions which adapt to the observed light curve. For extracting secondary frequencies, the dominant frequency is subsequently removed to analyse the resi-duals. The frequency range analysed was between 0.0003 to 0.025 cycles/day. Special attention has been paid to periods found near one year since a one-year peak alias appears in the spectral window of most of the stars. 
The appearance of candidate periods and aliases have been controlled adjusting the calculated combined light curve to the observations (see Fig. \ref{2pfit} and \ref{3pfit}).  Each individual frequency has been checked to reveal a consistent phased light curve of the residuals of remain dominant frequency.

\section{Results}

\begin{figure}
\centerline{\includegraphics[height=5cm,width=.4\textwidth,keepaspectratio]{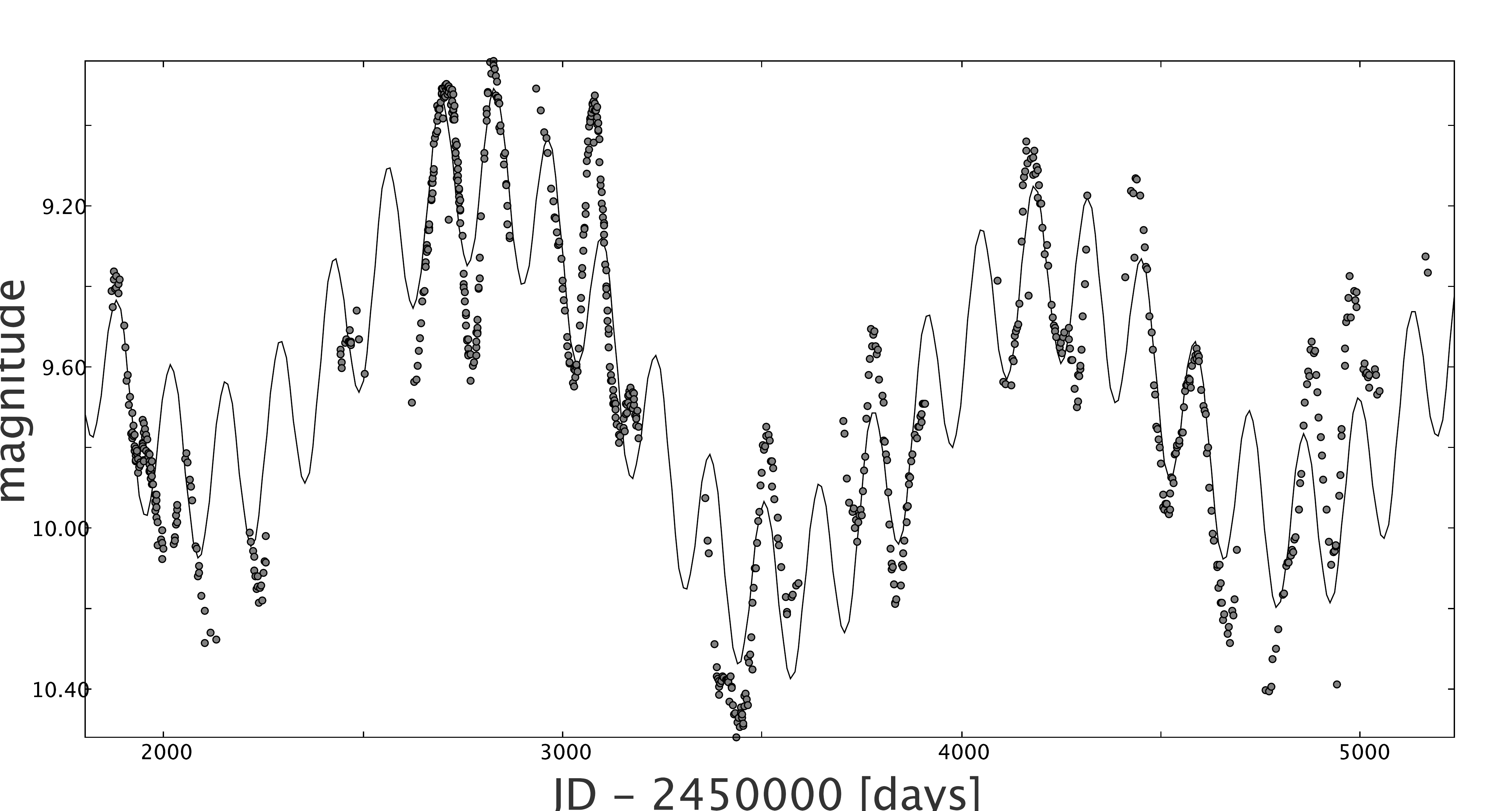}}
\centerline{\includegraphics[height=5cm,width=.4\textwidth,keepaspectratio]{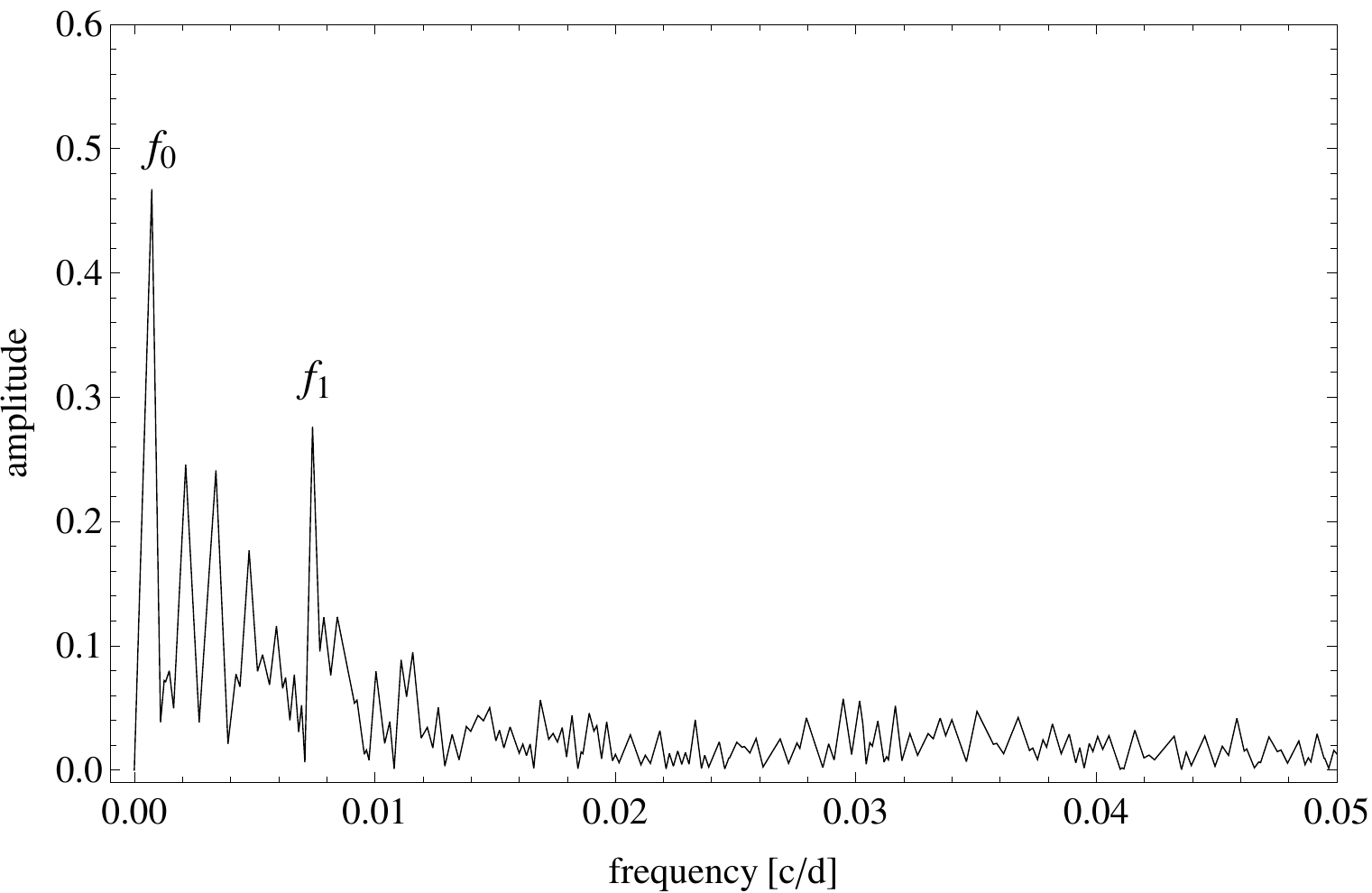}}

\caption{ Light curve and power spectrum of ASAS 112612-5121.6 showing 2 periods. The Fourier spectrum reveals two dominant peak frequencies that correspond to the  periods listed in table \ref{2P-tabla}.}
\label{2pfit}
\centerline{\includegraphics[height=5cm,width=.4\textwidth,keepaspectratio]{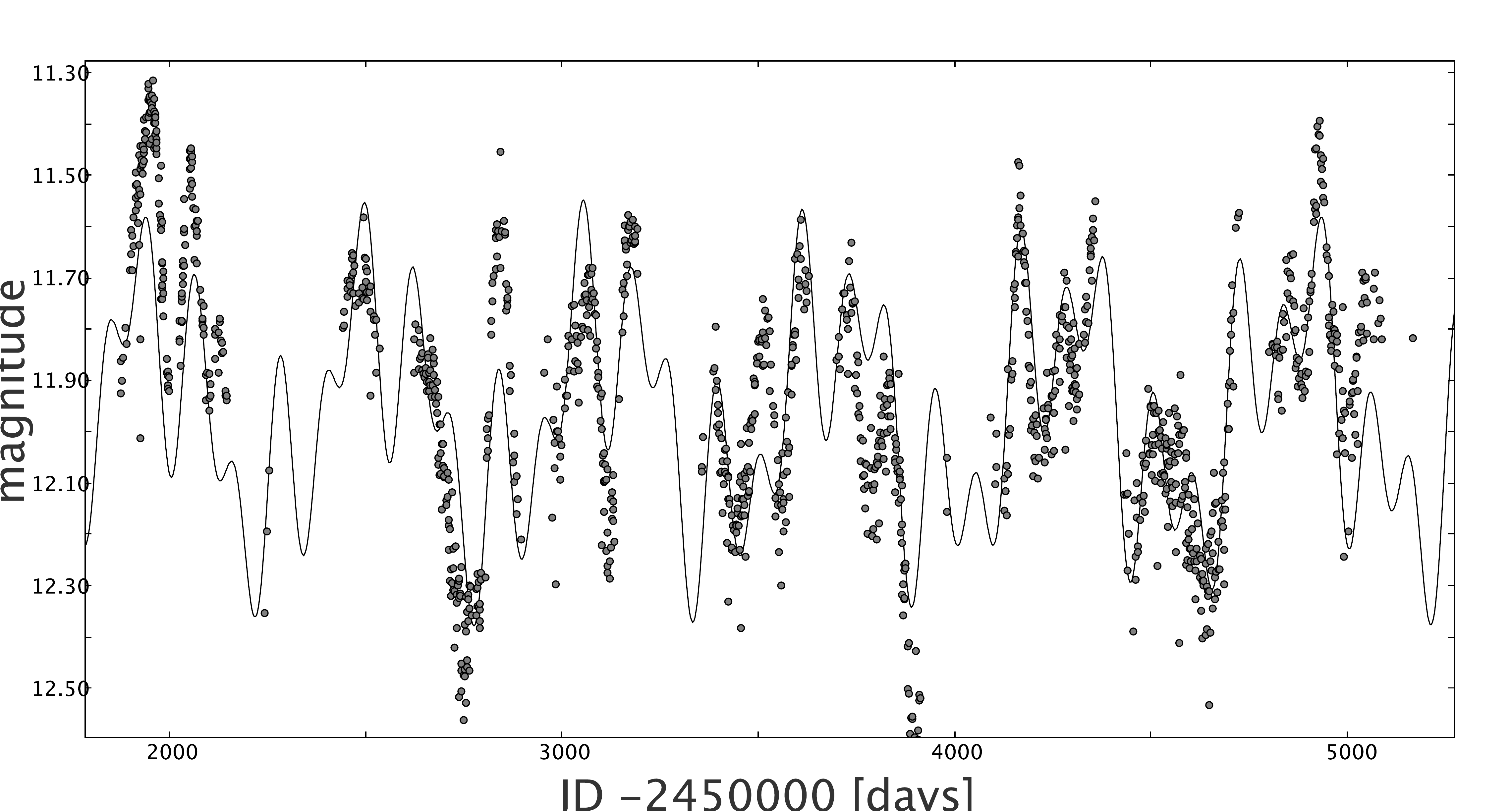}}
\centerline{\includegraphics[height=5cm,width=.4\textwidth,keepaspectratio]{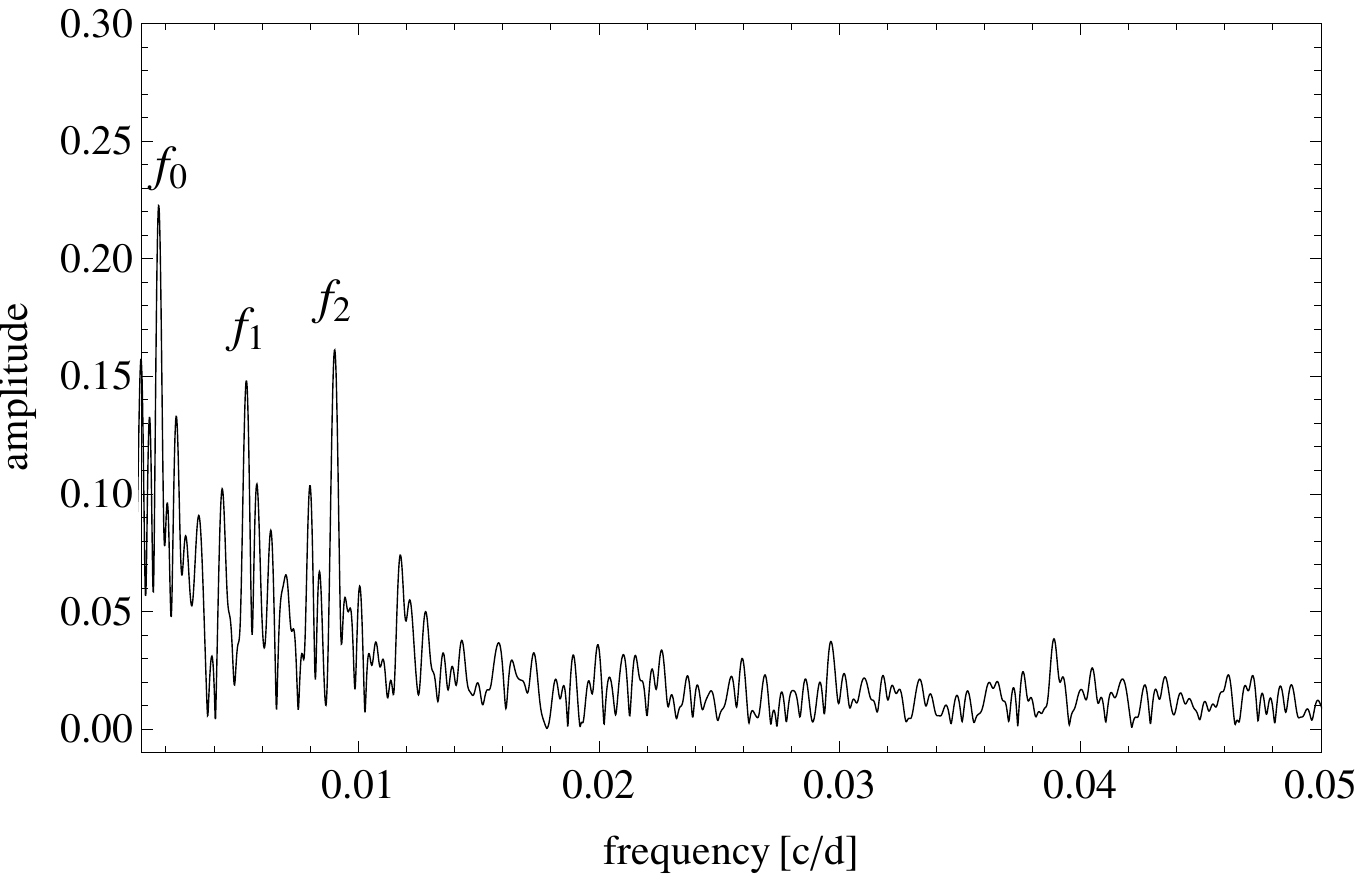}}
\caption{ Light curve and power spectrum of ASAS 131157-5757.8 showing 3 periods.
The Fourier spectrum reveals three dominant peak frequencies that correspond to the periods listed in table \ref{3P-tabla}.}
\label{3pfit}
\end{figure}

\begin{figure}
\centerline{\includegraphics[scale=0.82]{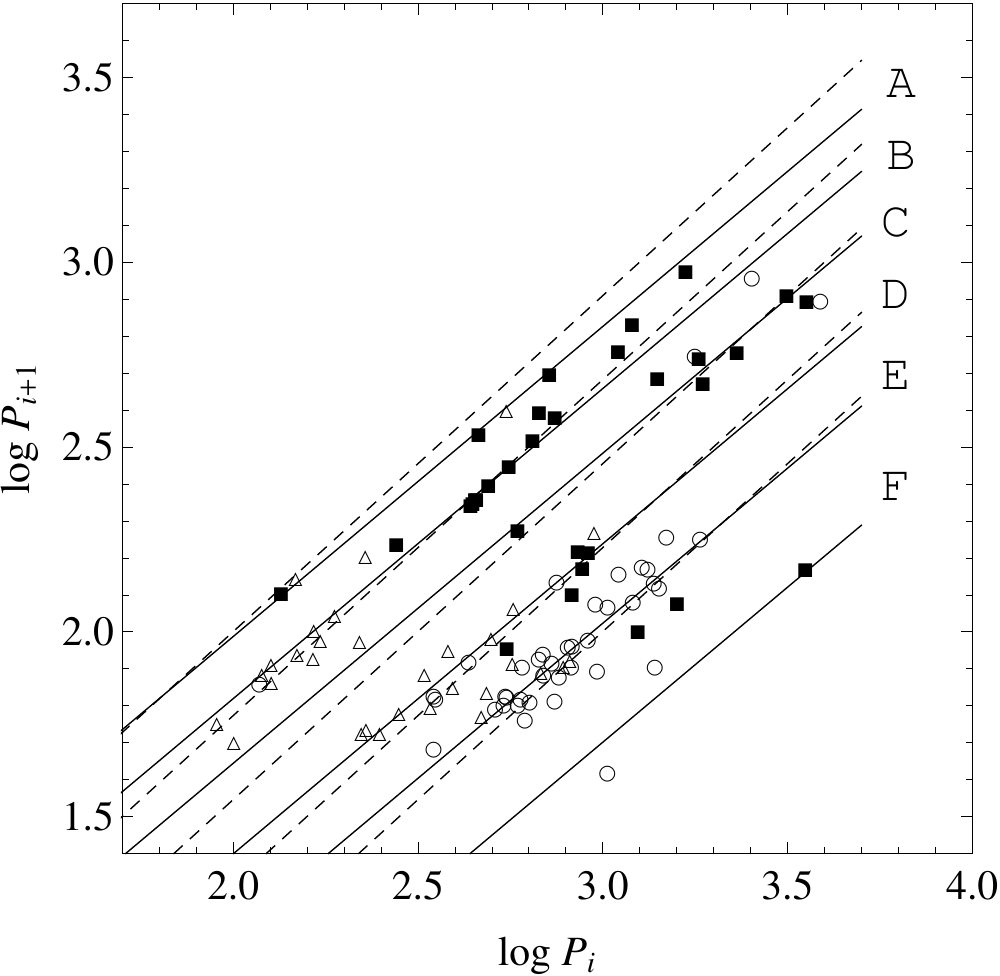}}
\caption{The shorter periods $\log P_{i+1}$ vs. the longer ones $\log P_{i}$ . Open circles  refer to  $i$ =$ 0$ for double periodic cases, filled squares to  $i$ =$ 0$ and  triangles to  $i$ = $1$ for triple periodic ones (which appear twice in the diagram). The qdashed lines refer to the locations of the sequences of Kiss et al (1999; their Fig. 8), the solid lines to linear least square fits on our data, adopting a mean value $\overline{d}_{2}$ = 0.839 for their slope.}
\label{Po-P1}
\vspace{3mm}
\centerline{\includegraphics[width=80mm,height=55mm]{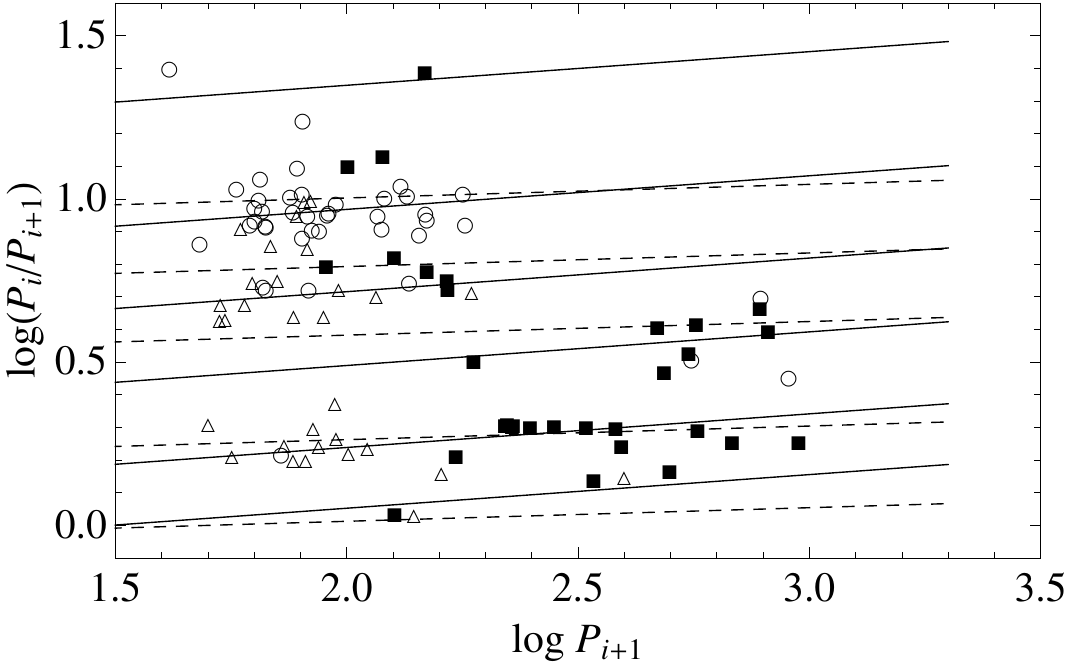}}
\caption{The Petersen diagram $\log (P_{i} /P_{i+1})$ vs. $\log P_{i+1}$ with the same distinction of symbols as in Fig. \ref{Po-P1}. The dashed lines refer to the locations of the sequences of Kiss et al (1999; their Fig. 9), the solid lines to linear least square fits on our data, adopting a mean value $\overline{p}_{2}$ = 0.103 for their slope. Sequence A corresponds to the bottom, F to the top of  the diagram.}
\label{petersen}
\end{figure}

Typical examples of light curves and their analysis are presented in Figures \ref{2pfit} and \ref{3pfit}, together with the corresponding power spectra. The periods and the semi-amplitudes determined by our analysis were listed in Table \ref{2P-tabla} for 41 stars with two periods, and in Table \ref{3P-tabla} for 31 stars with three periods. In addition, we list in these tables the classification of each star, according to Richards et al. (2012) \nocite{rich}. A total of 40 stars  (56\% of our sample) correspond to the class ``Long secondary period variables'' (LSP) as expected as dominant in our sample, because all these cases should present at least two different periods. Other 27 stars (37\%) are ``Semiregular pulsating variables'' (SRPV). The remaining 5 cases correspond to class RSG (red supergiants), RCB (R CrB stars) and Mira. \\
The double period diagram which compares adjacent periods (hereafter ``DP diagram'') is shown in  Fig. \ref{Po-P1}. This diagram (as well as that of Fig. \ref{petersen}) contains one point for each case with two periods, but two points for each case with three periods. These cases are distinguished by different symbols. Another presentation of the same data is given in Fig. \ref{petersen}, showing the so-called ``Petersen diagram'', which presents the period ratio vs. the shorter period. A similar diagram was first presented by Petersen (1973) in the context of an analysis of double mode Cepheids.\\
The locations of our sequences are very similar to those of Kiss et al. (1999)\nocite{kiss} which are also shown in Figs. \ref{Po-P1} and \ref{petersen}. Double and triple periodic stars are mixed in most sequences. We performed linear least square fits for each of the sequences in both diagrams:\\

\begin{equation}
\log{ P_{i+1}} = d_{1}+d_{2} \left( \log {P_{i} -1.7}\right) 
\label{logPo-P1}
\end{equation}

\begin{equation}
\log \left(P_{i}/P_{i+1}\right) = p_{1}+p_{2} \left(\log P_{i+1} -1.5\right)
\label{logPo/p1}
\end{equation}
\bigskip

\noindent
For double periodic cases $i$=$0$ is valid, while triple periodic cases both $i$=$0$ and $i$=$1$ have to be considered. The constant values in the right terms of equations (\ref{logPo-P1}) and (\ref{logPo/p1}) were chosen in a way that $d_{1}$ and $p_{1}$ correspond to the left axis intercept in Figs. \ref{Po-P1} and \ref{petersen}. Since the slopes $d_{2}$ and $p_{2}$ in all sequences turned out to be similar, we calculated mean values $\overline{d}_{2}$ = 0.839 and $\overline{p}_{2}$  = 0.103 averaging the individual slopes of all 6 sequences, weighted according to the number of data points N. Table \ref{tabla-fits} presents the corresponding coefficients $d_{1}$ and $p_{1}$ and the standard deviations $\sigma_{d}$  and  $\sigma_{p}$ for each sequence. The error bars of most of the data points in Figs. \ref{Po-P1} and \ref{petersen} are smaller than the symbol sizes in these figures. 
Therefore, the scatter present in these diagrams is real and not the result of uncertainties in the period determination. 

\begin{table}
\centering

\caption{Parameters of the sequences in Fig. \ref{Po-P1} and \ref{petersen}. N refers to the number of stars in each sequence (see text for details).}
\label{tabla-fits}
\begin{tabular}{cccccc}  
\hline \hline  
\\
&\multicolumn{2}{c}{Sequences in the }&\multicolumn{2}{c}{Sequences in the}& \\ &\multicolumn{2}{c}{DP diagram       }&\multicolumn{2}{c}{Petersen diagram}& \\
&\multicolumn{2}{c}{$\overline{d}_{2}$= 0.839 adopted}&\multicolumn{2}{c}{  $\overline{p}_{2}$= 0.103 adopted}\\
Sequence & $d_{1}$   & $\sigma_{d}$ & $p_{1}$   & $\sigma_{p}$ &N \\
\\
\hline \hline
 A      &1.735(5) & 0.007       & 0.001(9)    & 0.037    &  5\\
 B      &1.567(12)& 0.060       & 0.187(11)& 0.057    &  27\\
 C      &1.392(16)& 0.052       & 0.438(21)& 0.066    &  10\\
 D      &1.147(09)& 0.040       & 0.673(10)& 0.046    &  21\\
 E      &0.932(09)& 0.056       & 0.916(11)& 0.064    &  37\\
 F      &0.610(52)& 0.090       & 1.296(55)& 0.095    &  3\\       
 \hline
\end{tabular}
\end{table}

\section{Discussion and Conclusions}

The locations of our sequences in the DP diagram (Fig. \ref{Po-P1}) are very similar to those of Kiss et al. (1999)\nocite{kiss}, in spite of the fact that their sample was based on rather heterogeneous  amateur observations of very low accuracy (order of $\pm$ 0.3 mag), but covering long time intervals, several decades in some cases. Our data  refer to more homogeneous CCD observations covering only nine years with higher photometric accuracy (order $\pm$ 0.03 mag). There seems to be a systematic difference in the slopes: for our data it is smaller in Fig. \ref{Po-P1} and larger in Fig. \ref{petersen}, compared to Kiss et al. (1999)\nocite{kiss}. In addition, our sequence A is poorly populated (only 5 cases) which is a direct consequence of the way we selected the targets, by visual inspection of light curves. Since sequence A corresponds to two very similar periods, it is difficult to appreciate them directly in a light curve, compared to cases with larger period differences. On the other hand, we could populate well the sequences C and D, which  Kiss et al. (1999)\nocite{kiss} described as ``possible'' because they had only very few data points in this region of the DP diagram. Our data confirm these sequences. Our sequence F was not present in Kiss et al. (1999)\nocite{kiss}, but it has to be considered as uncertain because it contains only 3 data points with a rather large scatter. Finally, even the a bit larger position difference between sequence B and C in the Petersen diagram (Fig. \ref{petersen}), compared to the adjacent differences A$-$B and C$-$D, is clearly present in our data, as well as in those of Kiss et al. (1999)\nocite{kiss}. Existence and detailed positions of these sequences seem to be universal features, apparent in any data set of semiregular red giants of the AGB, which is only possible if they present different pulsation modes typical for similar stellar configurations. Other studies, often coming along from very different points of view, have leaded to similar conclusions. For instance, Soszynski et al. (2004)\nocite{sos} compared the Petersen diagrams of the LMC, SMC and Galactic bulge variable red giants, using OGLE-II data, and concluded that they are basically identical indicating that the variable red giants in all these different stellar environments share similar pulsation properties. Similar results were obtained by Tabur et al. (2010)\nocite{tabur}, who performed an extensive study of red giant branch pulsators in Milky Way, LMC and SMC; they concluded that the pulsation properties of red giants are consistent and universal, and propose to use these stars as high-precision distance indicators. \\
We understand our work as a pilot study, because it suffers still from the low number of cases. But we demonstrate that the ASAS data base is an excellent source for more extended investigations in this line. Semiregularily pulsating red giants are a very frequent class of stars, at least about 30\% of the variable stars in the ASAS data base are possible targets for such a study. This is underlined by the classification of Richards et al. (2012)\nocite{rich}; these authors list a total of 5096 stars of type LSP, which all should present more than one period, and 9982 stars of type SRPV which partly will contain multiperiodic variations, as demonstrated by our sample. This means that a systematic search for periodicities will reveal several thousands of multiperiodic cases, establishing finally the accurate positions of sequences in the DP and Petersen diagrams, implying this way a solid fundament for a complete understanding of the pulsation physics of these very numerous, but poorly understood  stars in the Universe. 

\acknowledgements
This research was supported by Grant DIUV 38/2011 of University of Valpara\'{i}so, Chile. 


\newpage
\end{document}